\documentclass[lettersize,journal]{IEEEtran}
\usepackage{amsmath,amsfonts}
\usepackage{algorithmic}
\usepackage{algorithm}
\usepackage{array}
\usepackage[caption=false,font=normalsize,labelfont=sf,textfont=sf]{subfig}
\usepackage{textcomp}
\usepackage{stfloats}
\usepackage{url}
\usepackage{verbatim}
\usepackage{graphicx}
\usepackage{cite}

\usepackage{hyperref}
\usepackage{cleveref}
\usepackage{multirow}
\usepackage{tabularx}
\usepackage{pifont}
\usepackage{xcolor}
\usepackage{listings}
\usepackage{fancyvrb} 
\usepackage{framed}

\usepackage{hyperref}
\hypersetup{
    colorlinks,
    linkcolor={blue!80!black},
    citecolor={blue!80!black},
    urlcolor={blue!80!black}
}

\usepackage{tikz}

\usepackage{tikz}
\newcommand*\emptycirc[1][1ex]{\tikz\draw (0,0) circle (#1);} 
\newcommand*\halfcirc[1][1ex]{%
	\begin{tikzpicture}
	\draw[fill] (0,0)-- (90:#1) arc (90:270:#1) -- cycle ;
	\draw (0,0) circle (#1);
	\end{tikzpicture}}
\newcommand*\fullcirc[1][1ex]{\tikz\fill (0,0) circle (#1);} 
\usepackage{orcidlink} 
\usepackage{booktabs} 
\usepackage{threeparttable} 

\newcommand{\bsub}[1]{\vspace{3px}{\noindent \textbf{#1}}}
\newcommand{\tsub}[1]{\vspace{2px}\noindent\textit{#1}}
\newcommand{\func}[1]{\textit{\texttt{#1}}}

\begin{document}

\title{OTA-Key: Over the Air Key Management for Flexible and Reliable IoT Device Provision}

\author{Qian Zhang$^{\orcidlink{0009-0003-9009-5946}}$,
        Yi He$^{\orcidlink{0000-0002-1807-4185}}$,
        Yue Xiao$^{\orcidlink{0009-0003-5047-5039}}$,
        Xiaoli Zhang$^{\orcidlink{0000-0002-5255-2216}}$,
        and Chunhua Song$^{\orcidlink{0009-0009-4404-4164}}$
\thanks{This paper was produced by the Taiyuan University of Technology, the Tsinghua University, the Wuhan University and the University of Science and Technology Beijing. \textit{(Corresponding authors: Chunhua Song and Xiaoli Zhang)}} 
\thanks{Qian Zhang and Chunhua Song are with College of Computer Science and Technology (College of Data Science), Taiyuan University of Technology, Taiyuan  030024, China (e-mail: zhangqian0452@link.tyut.edu.cn; songchunhua@tyut.edu.cn)}
\thanks{Yi He is with Department of Computer Science and Technology, Tsinghua University, Beijing 100084, China (e-mail: heyi21@mails.tsinghua.edu.cn)}
\thanks{Yue Xiao is with School of Cyber Science and Engineering, Wuhan University, Wuhan 430072, China (e-mail: yuexiao@whu.edu.cn)}
\thanks{Xiaoli Zhang is with School of Computer \& Communication Engineering, University of Science and Technology Beijing, Beijing 100083, China (e-mail: xiaoli.z@outlook.com)}
}



\maketitle

\begin{abstract}

As the Internet of Things (IoT) industry advances, the imperative to secure IoT devices has become increasingly critical. Current practices in both industry and academia advocate for the enhancement of device security through key installation. However, it has been observed that, in practice, IoT vendors frequently assign shared keys to batches of devices. This practice can expose devices to risks, such as data theft by attackers or large-scale Distributed Denial of Service (DDoS) attacks. To address this issue, our intuition is to assign a unique key to each device. Unfortunately, this strategy proves to be highly complex within the IoT context, as existing keys are typically hardcoded into the firmware, necessitating the creation of bespoke firmware for each device. Furthermore, correct pairing of device keys with their respective devices is crucial. Errors in this pairing process would incur substantial human and temporal resources to rectify and require extensive communication between IoT vendors, device manufacturers, and cloud platforms, leading to significant communication overhead. To overcome these challenges, we propose the OTA-Key scheme. This approach fundamentally decouples device keys from the firmware features stored in flash memory, utilizing an intermediary server to allocate unique device keys in two distinct stages and update keys. We conducted a formal security verification of our scheme using ProVerif and assessed its performance through a series of evaluations. The results demonstrate that our scheme is secure and effectively manages the large-scale distribution and updating of unique device keys. Additionally, it achieves significantly lower update times and data transfer volumes compared to other schemes.
\end{abstract}

\begin{IEEEkeywords}
Internet of Things (IoT), IoT security, Device, over the air (OTA), key update.
\end{IEEEkeywords}

\IEEEpeerreviewmaketitle

\section{Introduction}

\IEEEPARstart {W}{ith} the rapid evolution of modern communication technologies and the sweeping tide of intelligence across various sectors, an increasing number of smart devices are being deployed across different domains. These devices exchange data with cloud services, forming extensive Internet of Things (IoT) networks. Recent reports~\cite{DeviceNumbers,device_number2} forecast that by 2025, more than 10 billion devices will be interconnected with various cloud platforms, intensifying the challenges associated with securing these IoT devices.


A common security vulnerability in IoT devices is the tendency of vendors to use the same hardcoded keys, such as cryptographic credentials and secrets/passwords, across all their devices, as noted in \cite{share_keys, wu2024firmware, shared_iot}. 
When these shared keys are compromised, for instance, by extracting them from the firmware, attackers can jeopardize a vast number of devices. A notable example of this was the Mirai botnet, which compromised over 600,000 IoT devices in 2016 by exploiting known passwords \cite{mirai}. Furthermore, attackers can carry out large-scale device impersonation attacks \cite{iot_hazards}, where they masquerade as legitimate devices to connect to backend systems. This can lead to widespread device Denial-of-Service (DoS) attacks \cite{shared_iot} by manipulating the status of these devices.

\begin{figure*}[ht]
    \centering
    \includegraphics[width=\linewidth]{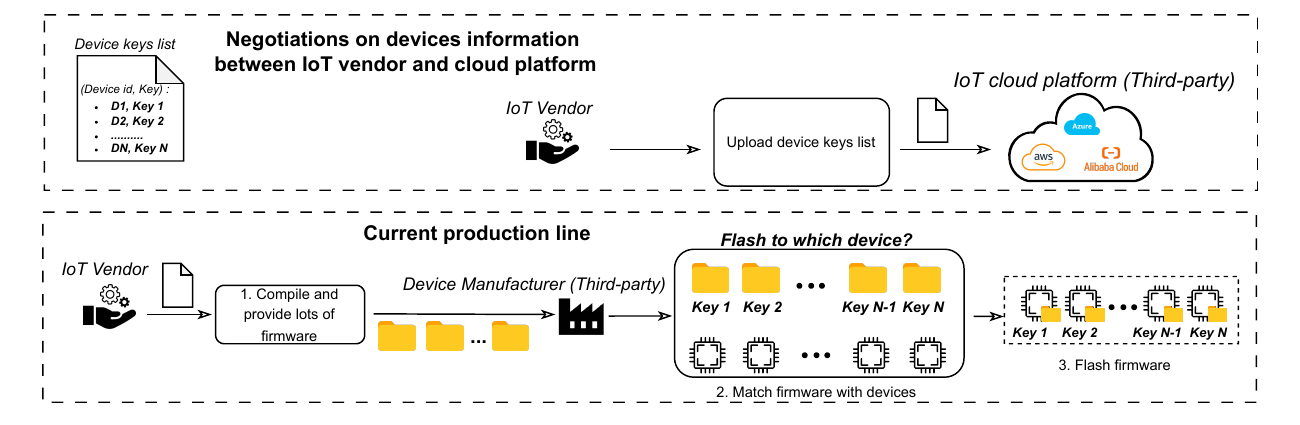}
    \caption{Current Production Line.} 
    \label{fig:system setting}
\end{figure*}

\bsub{Obstacles for Setting Unique Keys for Each Device:}
Setting unique keys for each device is challenging for device vendors. This challenge arises because IoT vendors typically outsource the manufacturing of their devices to device manufacturer, while provisioning these devices with unique keys requires multiple steps to manage and synchronize these keys between the factory and the cloud. As shown in Fig.\ref{fig:system setting}, in current production line, the IoT vendor needs to embed keys in the firmware of devices and register these keys to the device's backend server (i.e., IoT cloud). The obstacles are twofold. 
First, device keys must be correctly paired with the device IDs hardcoded on the chip, enabling the device to connect to the IoT cloud platform.
Any mismatch between device IDs and keys renders the IoT devices unusable, requiring an engineer more than \textit{20 minutes}~\cite{intel_sdo} to manually reflash the firmware to recover them. 
The reflashing process is time-consuming because it requires an engineer familiar with the firmware code \cite{rapidpatch} to physically retrieve the devices and then use a physical interface, such as ST-Link or JTAG, to flash the new firmware.
Second, the keys and device IDs must be synchronized with the cloud to ensure that the devices can be properly authenticated, increasing communication costs.
These two obstacles highlight the need for security and scalability, demanding that proposed solution addressing the issue of shared device keys should not alter the current production line.
Furthermore, we have identified that the current production line does not accommodate the need for updating device keys. Therefore, proposed solution also aims to fulfill the requirements.

\bsub{OTA-Key:} To fill these gaps, we intend to develop an agent-based device provisioning and management framework. 
It decouples device keys from the firmware and delegates key management to an independent agent.
The agent distributes and updates unique device keys on a large scale \emph{over the air}, eliminating the need for a wired connection (such as ST-Link or JTAG).
This process does not alter the existing production line while also supporting flexible updates of these unique keys. 

\bsub{Technical Challenges and Solution:} 
It is non-trivial to design and implement OTA-key. There are two technical challenges:

\tsub{Challenge 1: Distribute a unique key to each device without altering the existing industrial production line.} The existing industrial production process ( \cite{changhong, hisilicon, yiqismart}) employs two primary methods for distributing device keys. 
The first method involves flashing each device with different firmware.
After burning firmware, the device key is updated, then the firmware is recompiled and burned to the next device.
This process continues until all devices are successfully programmed. 
Although secure, this method is time-consuming and labor-intensive, making it impractical for large-scale device deployments. 
The second method involves sharing a common device key among a batch of devices by using the common firmware. 
While this approach is more efficient, it compromises security.
To achieve both scalability and security, we employ a two-stage device provisioning strategy to distribute unique keys to devices in a scalable fashion. Specifically, in the first stage, we burn the common firmware onto devices as the existing industry manufacturing does. In the second stage, when a device is powered on, the intermediary server (agent server) issues a unique key and platform connection information to the device.

\tsub{Challenge 2: Ensure seamless resumption of communication after network disruptions or power outages during updates.} In current industrial processes, it is not uncommon for devices to experience disruptions such as power outages or network disconnections during updates. Such incidents can lead to inconsistencies between the keys stored on the device and those maintained by the server, preventing the device from reconnecting to the server. The traditional recovery method requires returning the device to the factory for firmware reprogramming \cite{suzaki2020reboot}, which is labor-intensive and disrupts normal product usage. 
Our solution updates keys with an \textit{atomic agent-based device updating method}. This method allows devices that already have a unique key to securely acquire a new unique key through the agent.
We achieve seamless resumption by reserving a redundant area and implementing an update protocol, which ensures the old key is useful until the update process is complete.

\noindent\textbf{Results.} 
In our evaluation, we update device keys simultaneously in a scenario involving 1,000 devices. The process requires approximately 1,026 seconds and utilizes a data transfer volume significantly smaller than those of existing schemes, thereby illustrating the superior performance of our approach. Furthermore, we assess the overhead of this strategy on individual devices and the agent server, confirming its effectiveness and efficiency.

\noindent\textbf{Contributions.} The contributions of the paper are outlined as follows: \begin{itemize}
    \item \tsub{Scheme for distributing a unique key for each device.} We analyze the underlying reasons why IoT vendors often adopt a policy of using shared device keys in device production. In response, we develop OTA-Key, a secure, scalable, flexible, and reliable device provisioning and management scheme. This scheme enables the large-scale distribution of unique keys to devices and supports flexible updates.
    \item \tsub{Prototype implementation.} We implement prototypes of our scheme on both PC and real devices (specifically, the STM32F429I-DISC1 development board) for the server and device prototypes, respectively. 
    
    \item \tsub{Evaluation.}
    We use ProVerif to verify the security of the protocols in OTA-Key. The results confirm that our approach is secure. We also evaluate our approach based on device key update time and data transfer volume. Additionally, we assess the memory consumption of the server and the power usage of the devices. These evaluations demonstrate that our solution is highly efficient.
\end{itemize}

\section{Background And Motivation}

In this section, we first explain the workflow for setting unique keys for each device within the existing production line. Subsequently, we discuss the obstacles to achieving security within this workflow. Finally, we analyze the requirements that the workflow overlooks, and compare the functionalities of our scheme with existing approaches.

\begin{table*}
    \centering
    \begin{threeparttable}
        \caption{Functionality comparison of existing approaches.}
        \label{tab:func-comparison}
        \begin{tabular}{l|c|cccc}
            \toprule
             \multicolumn{2}{l|}{} & \textbf{Security} (Each device have a unique key) & \textbf{Scalability} & \textbf{Flexibility}  & \textbf{Reliability} \\
             \midrule
             \multirow{2}{*}{\textbf{Industrial practice}} 
            & Large Factory \cite{changhong, hisilicon} & \emptycirc & \fullcirc & \emptycirc & \emptycirc \\
            & Small Factory \cite{yiqismart} & \fullcirc & \emptycirc & \emptycirc & \emptycirc \\
            \midrule
            \multirow{5}{*}{\textbf{Academic literature}} 
           & Akkaoui et al. \cite{akkaoui_resilient_2024} & \emptycirc & \fullcirc & \emptycirc & \halfcirc \\
           & Asokan, N. et al. \cite{asokan_assured_2018} & \emptycirc & \fullcirc & \emptycirc & \halfcirc \\
           & Xu et al. \cite{xu_dominance_2019} & \emptycirc & \fullcirc & \emptycirc & \halfcirc \\
           & Wang et al. \cite{wang_design_2024} & \emptycirc & \fullcirc & \emptycirc & \halfcirc \\
           & Langiu et al. \cite{langiu_upkit_2019} & \emptycirc & \fullcirc & \emptycirc & \halfcirc \\
            \midrule
           \textbf{Our scheme:} & OTA-Key  & \fullcirc  & \fullcirc & \fullcirc & \fullcirc \\
            \bottomrule
        \end{tabular}
        \begin{tablenotes}
           \item  \fullcirc~Full support. \emptycirc~No support. \halfcirc~No specific implementation method has been provided.
        \end{tablenotes}
    \end{threeparttable}
\end{table*}

\subsection{Workflow for Setting Unique Keys to Each Device}
Before introducing the workflow, we first identify the five entities involved: IoT vendor, device manufacturer (DM), IoT cloud platform, device keys list and devices. Detailed descriptions of each entity follow:

\textit{\textbf{IoT vendor.}} 
IoT vendors are responsible for maintaining a large number of IoT devices and corresponding IoT services for end consumers. 
For example, IoT vendors like Bird~\cite{Bird}, Lime~\cite{Lime}, and DiDi~\cite{didi} provide micro-mobility services with shared electric bikes and scooters. In practice, an IoT vendor needs to select a suitable device manufacturer (DM) to produce IoT devices and choose an appropriate IoT cloud platform to deploy IoT services.


\textit{\textbf{Devices manufacturer.}} 
The device manufacturer (DM), such as~\cite{changhong}, produces IoT devices and flashes firmware compiled by IoT vendors. Typically, the firmware includes the connection addresses of the IoT cloud platform and keys (such as X.509 certificates and secrets/passwords) to ensure secure connections between IoT devices and the cloud platform.

\textit{\textbf{IoT cloud platform.}} 
IoT cloud platforms, such as AWS \cite{awsDeviceProvision}, Azure \cite{azureDeviceProvision}, and Alibaba Cloud  \cite{aliyundeviceprovision}, provide various services that include collecting data from IoT devices and enabling control over these devices to execute commands.

\textit{\textbf{Device keys list.}} The device keys list contains a mapping of device IDs to their respective keys and is managed and generated by the IoT vendor.

\textit{\textbf{Devices.}} 
IoT devices, such as gateways, network surveillance cameras, and shared e-scooters. 

\noindent\textbf{Workflow.} The workflow for setting unique keys (such as X.509 certificates and secrets/passwords) to devices in the current production line, as illustrated in Fig.~\ref{fig:system setting}, is as follows:
\begin{enumerate}
    \item IoT vendors compile a unique key for each firmware, then the firmware is delivered to the Device Manufacturer (DM).
    \item The DM matches the firmware with the devices, which is error-prone (analyzed in \S~\ref{subsec:Match is error-prone.}). 
    \item The DM flashes the firmware to the devices.
\end{enumerate}

The IoT vendor is required to upload the devices key list, which includes device IDs and corresponding keys, to the cloud platform.

\subsection{Obstacles for Setting Unique Key for Each Device}
\label{subsec:Match is error-prone.}
\noindent\textbf{Matching firmware with device is error-prone.}
It is necessary to match the device key within the firmware with the ID embedded in the device's chip.
However, the ID must be read from the device itself, while the firmware containing the key is provided by the IoT vendor.
Consequently, the device IDs and the device keys, generated by different entities, complicate the matching process.
Additionally, flashing firmware onto devices on a large scale is both time-consuming and labor-intensive, due to the sheer volume of devices involved and the inability to distinguish IDs and keys by merely looking at the devices or firmware.
According to \cite{intel_sdo}, modifying and flashing firmware under the existing production line conditions requires an engineer more than 20 minutes per device. These factors significantly increase the likelihood of errors during the matching process. 

\textit{Why must the ID be read from device?} Current industrial practice involves using a common product key for all devices to request IDs and tokens from a server. This method is insecure as the server cannot verify whether a device has already obtained a key, potentially enabling abuse by attackers. Additionally, the ability for one ID to request multiple tokens can be exploited to knock legitimate devices offline.

\noindent\textbf{High communication costs.} IoT vendors often negotiate with Device Manufacturers (DMs) and multiple cloud platforms, increasing costs due to collaborations across various entities.



\subsection{The Need for Flexible Key Management}
\label{subsec:need-flexible-key}
In the current product line, IoT vendors do not account for the need to update device keys to address the issue of key leakage, as highlighted by \cite{mosenia_comprehensive_2017, el_jaouhari_secure_2022}. To update device keys within the existing production process, the device must be returned to the Device Manufacturer (DM) for firmware re-burning, which incurs substantial costs. Additionally, the platforms themselves are vulnerable to attacks, as noted by \cite{ruleder}, further necessitating flexible key updates by vendors. Moreover, according to \cite{draft_swithCloud}, a recent EU draft mandates that IoT devices be capable of easily switching cloud services to prevent data monopolization by large cloud platforms. This regulation underscores the need for updating device keys to enhance flexibility and security.

We conduct an extensive survey of industrial practices and academic literature, summarizing the results in Table~\ref{tab:func-comparison}. This table shows that existing methods achieve either security or scalability; however, none provide the flexibility required to switch cloud platforms.

\section{Threat Model and Design Overview}

\begin{figure*}[ht]
    \centering
    \includegraphics[width=\linewidth]{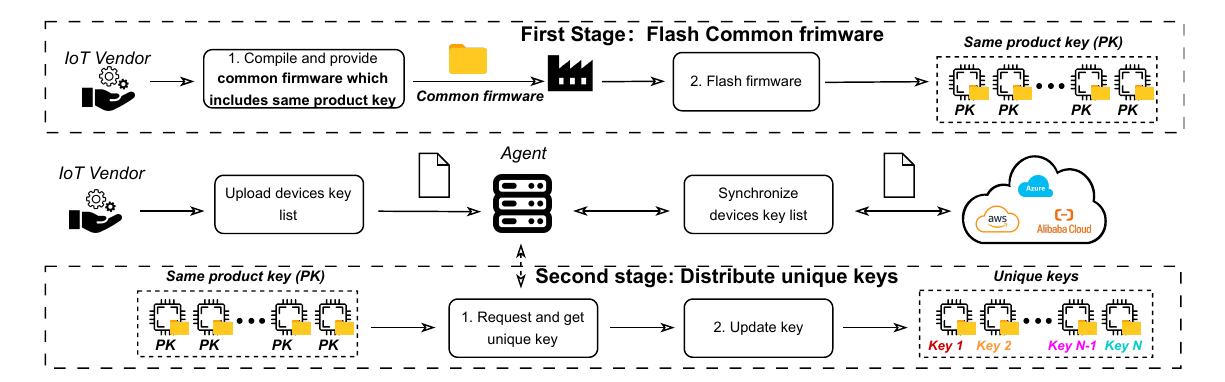}
    \caption{Our Design.} 
    \label{fig:initial workflow}
\end{figure*}

\subsection{Threat Model} We assume that attackers aim to launch large-scale attacks to compromise the availability and integrity of IoT devices. In practice, we assume they can access a limited number of device keys through reverse engineering. 

However, this attack capability is constrained by the time-consuming nature of reverse engineering~\cite{wu2024your, gustafson2019toward} and the risk of exposure~\cite{dauterman2020safetypin}.
IoT vendors are considered trusted, while manufacturers are categorized as honest-but-curious. Manufacturers may attempt to record device keys and collude with external attackers to compromise lots of IoT devices.
The cloud is also honest-but-curious. Here, we assume privacy-preserving IoT services are deployed via techniques like Trusted Execution Environments (TEEs)~\cite{intelsgx}, which ensures that the keys of IoT devices remain secure on the cloud.

\textit{Hazards.}
The primary objective of an attacker, under this assumption, is to obtain the keys of all devices, and then control them or impersonate legitimate devices.
For instance, the Mirai botnet attack \cite{mirai} controlled over 600,000 compromised IoT devices by exploiting their shared keys and launched DoS attacks against numerous mainstream websites.

\subsection{Design Goal}
\label{subsec:design-goal}
Our device provisioning scheme (OTA-Key) should achieve the following design goals.
\begin{itemize}
    \item \textbf{Security:} Our scheme ensures that even if some device keys are leaked or the firmware-burning process is exploited by manufacturers, the compromise is limited and cannot lead to the exposure of keys for a large number of devices, thereby preventing the escalation to a large-scale attack (e.g., Denial of Service~\cite{shared_iot}).
    
    \item \textbf{Scalability:} The entire process of device provisioning and management can be easily applied to large-scale IoT deployments without modifying the existing production line.
    
    \item \textbf{Flexibility:} 
    IoT devices should have the capability to flexibly switch between cloud platforms and update device keys (such as X.509 certificates and secrets/passwords) for several reasons \cite{ kazim_framework_2018, cao_scalable_2020}.

    \item \textbf{Reliability:} Key information in IoT devices should be updated reliably, as IoT devices are constrained devices (e.g., with limited processing power, memory, and battery life) and are often in environments with unstable network conditions. 
    
\end{itemize}

\subsection{Design Overview} 

We propose an agent-based device provisioning and management framework, where we abstract the large-scale device key management away from the firmware and manage them independently.
Specifically, the designs related to per-device keys are abstracted away and delegated to \emph{an agent}.
Other designs relevant to firmware are generalized and completed by the existing production lines of IoT vendors and device manufacturers.
Such a framework ensures security without altering the existing production lines and updates device keys \emph{over the air} to support large-scale deployments.
The agent helps provision and update devices via the following two key designs.

\noindent \textbf{Two-stage device provisioning strategy.} 
As illustrated in Fig.~\ref{fig:initial workflow}, we devise a \emph{two-stage device provisioning strategy} to assign unique keys to devices in a scalable manner.
Specifically, in the first stage, we burn the common firmware which includes the same product key (PK) onto devices as existing industry manufacturing practices.
In the second stage, when a device is powered on, the agent issues a unique key and platform connection information to the device. In this way, we can assign a unique key to each device. More importantly, since the unique keys are assigned after the firmware is burned, the curious device manufacturer will not get keys (achieving {the security goal}). Moreover, since the first stage is a part of the existing production line and the second stage is another independent process, we will not alert the existing production line, thereby meeting the scalability goal.

\noindent \textbf{Atomic agent-based device updating method.}
After device provisioning, the device can communicate with the cloud platform using the unique key and connection information set by the agent.
Once the cloud platform or the key is required to be updated due to security concerns or production plan changes, the \emph{atomic agent-based device updating method} will be utilized.
Devices can update their keys and cloud information through the agent.
The update process follows an atomic protocol, which ensures the correctness of the update even facing power outages or unstable network conditions, thereby meeting the reliability goal. 
Additionally, the only requirement for obtaining the key and connection information in the device is to connect to the agent, without the need for complex operations such as registering for a new key from the cloud, or reburning the firmware.
And the method can be used to switch cloud platforms, achieving the flexibility goal.

\section{Design Details}

In this section, we introduce the design details, including the two-stage device provisioning strategy and the atomic agent-based device updating method. Related abbreviations can be found in Table \ref{tab:Notation Summary}.

\begin{table}[ht]
    \caption{Notations}
    \centering
    \renewcommand{\arraystretch}{1.5}
    \begin{tabular}{p{3cm} p{4cm}}
    \hline
      Notations & Descriptions \\
      \hline
      \textit{ID} & device id\\
      \textit{PO} & product order \\
      \textit{PK}& product key\\
      \textit{AK}& agent key, used for communication between device and agent\\
      \textit{Nonce} & a random number\\
      \textit{Enc(K, M)} & the ciphertext of \textit{M} encrypted with the key \textit{K} \\
      \textit{HMAC(K, M)} & hash-based message authentication code of \textit{M} with the key \textit{K} \\
      \hline
    \end{tabular}
    \label{tab:Notation Summary}
\end{table}

\subsection{Two-stage Device Provisioning Strategy}
\label{subsec:two-stage}

\begin{figure}
    \centering
    \includegraphics[width=0.5\textwidth]{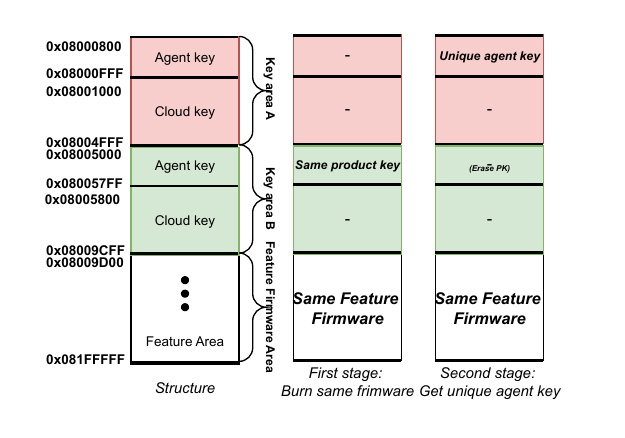}
    \caption{Flash memory structure design. }
    \label{fig:flashStruct}
\end{figure}

The strategy involves two stages. In the first stage, we burn the common firmware. And in the second stage, we distribute the unique device key. A unique device key contains two keys: the agent key, which is used to communicate with the agent, and the cloud key, which is used to connect to the cloud platform. The entire processing relies on the flash memory structure. Below, we explain this design and show the initialization of unique device keys.

\noindent\textbf{Flash memory structure.} We partition the firmware based on key and feature functionality, as illustrated in Fig.~\ref{fig:flashStruct}. (Structure), we allocate the initial 18KB of storage, known as key area A, which is further divided into two segments: the first 2KB, ranging from \textit{0x08000800} to \textit{0x08000FFF}, is dedicated to storing the unique agent key; the subsequent 16KB, from \textit{0x08001000} to \textit{0x08004FFF}, stores the cloud key and cloud connection addresses. Following area A, an additional 18KB (from \textit{0x08005000} to \textit{0x08009CFF}), designated as area B, is reserved as redundancy space to support atomic updates of keys. The remaining portion of the flash memory is used to store the feature firmware, which is identical across devices within the same batch. (`First stage' and `Second stage') in the first stage, DM programs a batch of devices with the same product key in key area B and the same feature firmware. In the second stage, IoT vendors enable devices to initiate a request to the agent, receive a unique agent key, and erase the same product key. The STM32F4 series designates \textit{0x08000000} as the start address for the flash memory. The 2KB space from \textit{0x08000000} to \textit{0x080007FF} is reserved for the interrupt vector table. Consequently, our design for the flash memory structure begins at \textit{0x08000800}.

\noindent\textbf{Initialization of unique device keys.}
In the first stage, a same product key is burned to all devices.
In the second stage, a device uses this product key to initiate a request for a unique agent key from the agent. The new agent key is then stored on the device, and the product key is discarded. Next, the device uses the new agent key to communicate with the agent and obtain the cloud key.

\begin{figure}
    \centering
    \includegraphics[width=0.48\textwidth]{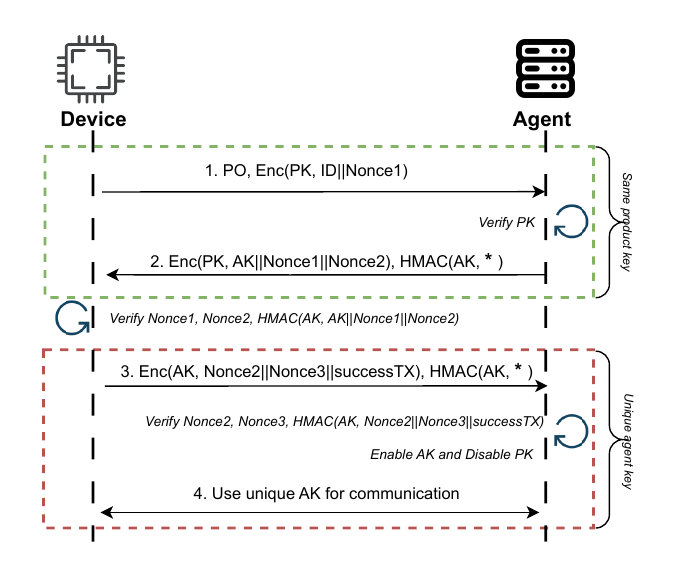}
    \caption{Agent key initialization procedure.}
    \label{fig:get-DS}
\end{figure}

\textit{Initialization of Unique Agent Keys.}\label{Ini:unique agent key}
Fig.~\ref{fig:get-DS} illustrates the detailed process through which a device acquires the agent key. 
The device ID, stored within the chip, can be exemplified by the STM32F410, where it is accessed from the base address \textit{0x1FFF7A10} (96 bits of data) using the same firmware \cite{uid_stm32_manual}. Additionally, unique keys can be generated using random number generators (RNG), with various RNG technologies described in \cite{sunar_provably_2007, chugunkov_classification_2017}. The process of obtaining the agent key is specifically outlined as follows:
\begin{enumerate}
    \item The device sends a product order (\textit{PO}) to the agent, along with the ciphertext of the device id (\textit{ID}) and a random number (\textit{Nonce1}) encrypted with the product key (\textit{PK}), formatted as \textit{Enc(PK, ID$\vert$$\vert$Nonce1)}.
    \item The agent decrypts the message using the PK corresponding to the \textit{PO}; if the decryption is successful, it retrieves the \textit{ID}. Subsequently, the agent sends \textit{Enc(PK, AK$\vert$$\vert$Nonce1$\vert$$\vert$Nonce2)}, along with \textit{HMAC(AK, *)}, where `\textit{*}' represents `\textit{AK$\vert$$\vert$Nonce1$\vert$$\vert$Nonce2}'.
    \item The device decrypts the information using \textit{PK} to obtain the unique agent key and verifies the \textit{HMAC}. Then, the device sends \textit{Enc(AK, Nonce2$\vert$$\vert$Nonce3$\vert$$\vert$successTX)}, along with the corresponding \textit{HMAC}. The purpose of \textit{successTX} is to inform the agent that the agent key has been updated.
    \item The agent verifies the \textit{Nonce} and \textit{HMAC}. Once verified, the device and agent can communicate using the unique agent key.
\end{enumerate}

The product key (\textit{PK}) is pre-installed on both the device and the agent.
Before obtaining the unique agent key, the device communicates with the agent using the \textit{PK}.
This approach ensures that malicious devices without the \textit{PK} cannot initiate a valid request, and malicious agent without the \textit{PK} cannot send a false agent key.
The unique agent key can be securely allocated. Then the agent provides unique cloud keys to devices.

\begin{algorithm}
\caption{Unique Agent Key Requesting Algorithm}

\textbf{Input:} device, request

\textbf{Output:} AK (if conditions met)
\label{alg:requestAK}
\begin{algorithmic}[1]
\STATE PO $\gets$ \func{GetProductOrder}(device)
\STATE ID $\gets$ \func{GetDeviceID}(device)
\STATE PK $\gets$ \func{GetProductKey}(device)
\STATE Nonce1 $\gets$ \func{GetNonce}(request)
\STATE payload $\gets$ \func{Concat}(ID, Nonce1)
\STATE msg $\gets$ \func{Concat}(PO, \func{ENC}(PK, payload))
\STATE \func{SendToAgent}(msg, request)
\STATE responseMsg $\gets$ \func{GetResponse}(request)
\STATE dec\_responseMsg $\gets$ \func{DEC}(PK, responseMsg)
\STATE recived\_Nonce1 $\gets$ \func{Parse}(dec\_responseMsg)
\STATE Nonce2 $\gets$ \func{Parse}(dec\_responseMsg)
\STATE AK $\gets$ \func{Parse}(dec\_responseMsg)
\STATE recived\_HMAC $\gets$ \func{GetHMAC}(responseMsg)
\IF{Nonce1 = recived\_Nonce1}
    \IF{Nonce2 is new}
        \IF{recived\_HMAC = \func{HMAC}(AK, \texttt{*})}
            \STATE Accept AK
        \ELSE
            \STATE Exit
        \ENDIF
    \ELSE
        \STATE Exit
    \ENDIF
\ELSE
    \STATE Exit
\ENDIF
\STATE \textbf{return} AK
\end{algorithmic}
\end{algorithm}

To further clarify the process of obtaining the agent key, we have developed a unique agent key requesting algorithm, where the device requests the agent key from the agent. As shown in Algorithm \ref{alg:requestAK}, the device retrieves PO, ID, and PK (lines 1-3), generates a random number Nonce1, and encrypts the ID and Nonce1 using PK to create a message (lines 4-6). This message is sent to the agent (line 7). Upon receiving the response containing the agent key (AK) encrypted with PK, the device decrypts the AK (line 8-13) and validates the response by checking the Nonce1 match, ensuring Nonce2 is unused to prevent replay attacks, and verifying the HMAC value for message integrity. After successful validation (line 14-16), the device accepts the agent key.

\textit{Initialization of Cloud Keys.} The process of initializing the cloud key is depicted in Fig.~\ref{fig:get-connection}. Since the initialization of the cloud key can be considered a special update process, it will be detailed in the following subsection.

\noindent{\textbf{Remark.}} The above device provisioning process is secure and scalable. 

{\emph{1) Security:}}
We issue a unique agent key to each device, eliminating the shortcomings associated with sharing the same key between devices. Device keys are assigned to devices online and after the firmware is burned, ensuring that the device manufacturer does not learn them. Moreover, the entire process can be completed in a secure environment; for example, the first stage can be done within the industry, and the second stage can be completed before the devices are released to the market.
{\emph{2) Scalability:}}
Since the firmware burned into each device during the first stage is identical, this process occurs without altering the existing industrial production process. In large-scale production, the IoT vendor only needs to prepare one version of the firmware. Additionally, the second stage is lightweight and efficient, even when managing a large number of devices.

\subsection{Atomic Agent-Based Device Updating Method}
\label{sec:atomocUpdate}

After distributing unique device keys and deploying devices in the market,
there are two main reasons for updating device keys: first, regularly updating the agent key enhances the security of the device; second, devices may need to change the cloud platform. However, the update process can be disrupted for several reasons, particularly because IoT devices are constrained by limited processing power, memory, and battery life, and they frequently operate in environments with unstable network conditions or where power interruptions may occur. Therefore, we design the \emph{atomic agent-based device updating method} to provide flexible and reliable updates. This method updates device keys flexibly via the agent and ensures reliability by reserving a redundant area.

\noindent \textbf{Update protocol for unique agent keys.}
The process of updating the agent key follows a procedure similar to the \emph{agent key initialization procedure} (\S~\ref{Ini:unique agent key}~Fig.~\ref{fig:get-DS}), and thus the details of the update process are not reiterated here. The difference between the two processes lies in step 1, where the device requests a new agent key (\textit{AK}) from the agent using the old \textit{AK} instead of the product key (\textit{PK}).

\begin{figure}
    \centering
    \includegraphics[width=0.5\textwidth]{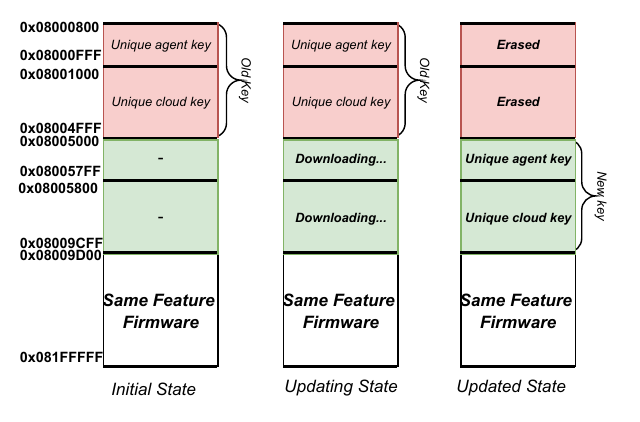}
    \caption{Flash memory state during key update. }
    \label{fig:A/B Scheme update}
\end{figure}

\noindent \textbf{Update protocol for unique cloud keys.}
Fig.~\ref{fig:get-connection} depicts the detailed process by which a device acquires or updates its cloud key and connection information.
In short, the device authenticates to the agent using its agent key, after which the agent registers the device with the cloud to obtain the cloud key and connection information, and finally the agent transmits them to the device.
Fig.~\ref{fig:A/B Scheme update} illustrates changes in the device's flash memory during the update process. The fundamental concept involves writing new key into a redundant key area designated for key storage. During this process, the device's flash memory transitions through three distinct states: the Initial State, where the old key remains active and the redundant key area is unused; the Updating State, where the new key is being written into the redundant area; and the Updated State, where the new key is activated, and the old key has been erased, preparing the memory for future updates.
Once the new key is written and verified, the new key is activated and the one is erased. This erased portion will be used to store keys for future updates.

The specific steps are as follows:

\begin{figure}
    \centering
    \includegraphics[width=0.5\textwidth]{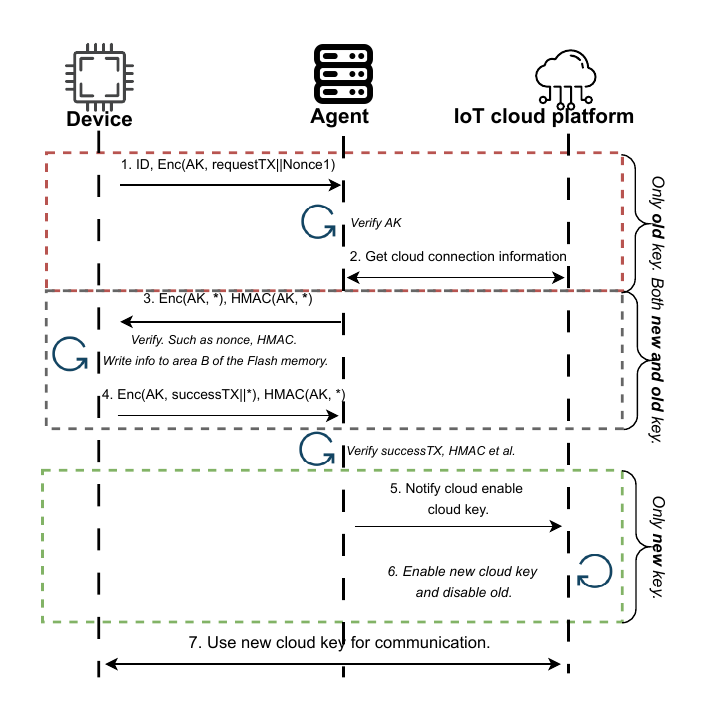}
    \caption{Cloud key update procedure.} 
    \label{fig:get-connection}
\end{figure}

\begin{enumerate}
    \item The device sends its \textit{ID} and \textit{Enc(AK, requestTX$\vert$$\vert$ Nonce1)} to the agent. The \textit{requestTX} indicates the device’s request is aim to obtain the cloud key and connection information.
    \item The agent verifies the \textit{AK}. Subsequently, the agent retrieves connection information from the cloud platform. Before the completion of step 2, the state of the device's flash corresponds to the leftmost diagram \textit{(Initial State)} in Fig.~\ref{fig:A/B Scheme update}.
    \item The agent sends \textit{Enc(AK, *)}, \textit{HMAC(AK, *)}, where `\textit{*}' represents the cloud key and connection information.
    \item The device verifies the \textit{Nonce} and \textit{HMAC}. Once verified, the device writes the information into area B of the flash memory. And the device sends \textit{Enc(AK, successTX$\vert$$\vert$*)}, \textit{HMAC(AK, *)}, where the purpose of \textit{successTX} is to inform the agent that the cloud key and connection information has been updated, and `\textit{*}' represents other necessary related information. Steps 3-4 involve the downloading and verification of the new key. Before the completion of step 4, the state of the flash is depicted in the middle diagram \textit{(Updating State)} in Fig.~\ref{fig:A/B Scheme update}.
    \item The agent verifies the \textit{HMAC} and \textit{successTX}, and notifies the cloud to activate the new cloud key and connection information and disable the old ones.
    \item The cloud, upon receiving the agent’s notification, activates the new cloud key and connection information and disables the old ones. Steps 5-6 involve the agent sending commands to the cloud to enable the new key and disable the old key after the key update is completed. Between the completion of step 4 and the initiation of step 5, the state of the device's flash corresponds to the rightmost diagram \textit{(Updated State)} in Fig.~\ref{fig:A/B Scheme update}.
    \item The device communicates with the cloud using the new cloud key and connection information.
\end{enumerate}


This approach achieves atomic updates of the cloud key while ensuring that the key stored between the device and the cloud remains consistent. 
Before the new key begins to be written to the device (steps 1 to 2), the cloud platform only supports connections with the old key; while the agent is writing the new key to the device and awaiting a confirmation message (steps 3 to 4), the cloud platform supports both the old and new keys for connections. After the agent receives the confirmation from the device (after step 5), the cloud platform exclusively supports connections with the new key. The flash structure ensures that the old key is not overwritten during the key update process, thereby achieving atomic updates.

\lstset{frame=single, caption={Formal Verification Code}, label={lst:formal verification},float=ht}
\begin{lstlisting}
(* Types *)
type key.

(* Declare free variables *)
free PK: key [private].
free AK: key [private].
free ID: bitstring.
free successTX: bitstring [private].

(* Declare private channels *)
free c : channel.

(* Functions and Constructors *)
fun enc(key, bitstring): bitstring.
reduc forall m: bitstring, k: key; dec(k,enc(k,m))=m.

fun hmac(key, bitstring): bitstring.
fun pair(bitstring, bitstring): bitstring.
fun keyToString(key): bitstring.

(* Events *)
event begin_init(bitstring, bitstring).
event end_init(bitstring, bitstring).

(* Protocol definition *)
let protocol(ID: bitstring, Nonce1: bitstring) = 
  (* Step 1: Device -> Agent : Enc(PK, ID || Nonce1) *)
  out(c, enc(PK, pair(ID, Nonce1)));
  event begin_init(ID, Nonce1);

  (* Step 2: Agent -> Device *)
  new Nonce2: bitstring;
  in(c, recived_msg:bitstring);
  let dec_recived_msg=dec(PK, recived_msg) in
  let mac2 = hmac(AK, dec_recived_msg) in
  out(c, enc(PK, pair(keyToString(AK), dec_recived_msg)));
  out(c, mac2);

  (* Step 3: Device -> Agent *)
  new Nonce3: bitstring;
  in(c, recived_msg2:bitstring);
  let dec_recived_msg2=dec(PK, recived_msg2) in
  let mac3 = hmac(AK, dec_recived_msg2) in
  out(c, enc(AK, pair(Nonce2, dec_recived_msg2)));
  out(c, mac3);

  (* Step 4: Use unique AK for communication *)
  in(c,recived_msg3: bitstring);
  let dec_recived_msg3=dec(AK, recived_msg3) in
  out(c, enc(AK, pair(successTX, dec_recived_msg3)));
  event end_init(ID, Nonce1).

(* Ensure PK and AK are kept secret *)
query attacker(PK).
query attacker(AK).
query attacker(successTX).

(* Run the protocol *)
process 
  new Nonce1: bitstring;
  !protocol(ID, Nonce1)
\end{lstlisting}

Furthermore, we analyze two scenarios to illustrate how this protocol ensures reliable updates for devices. In the first scenario, if a power outage occurs and disrupts step-3, steps 4-7 will not proceed, and the device does not acquire the complete new key, leaving the old key unchanged. Meanwhile, since the cloud has not received instructions to activate the new key, it continues to support connections via the old key. Therefore, upon power restoration, the device can resume normal communication with the cloud using the old key. 
In the second scenario, if the agent fails to receive confirmation (step-4) from the device, but the device has already begun using the new key for connections, this action would prompt the cloud to disable the old key and enable the new one.

\noindent \textbf{Remark.}
The device updating method achieves flexibility and reliability:
\textit{1) Flexibility:}
Only the keys need to be changed in the process, without updating the entire firmware. If devices need to change the cloud platform and the firmware should be updated accordingly, the cloud platform can prepare only one firmware, and the key update will occur after the firmware update (further discussed in \S~\ref{discussion}).
\textit{2) Reliability:}
The designed protocol can address problems that arise on the device side, such as device power-offs. Additionally, the existing protocol (such as TCP/IP) ensures the reliability of remote communication. In this manner, a new key will only be used after both the agent and the device have confirmed it. Furthermore, the flash memory structure ensures that a failed update does not affect the common use of devices.

\section{Evaluation}

In this section, we evaluate our design. We utilized ProVerif to verify the security of our proposed scheme in \S~\ref{chp:eva-formal} (achieving security goal) and present the prototype implementation in \S~\ref{chp:eva-impl}. We then compare the performance of our approach with other update baselines and assess the impact of firmware size on key update times for a single device, detailed in \S~\ref{chp:eva-update-time}. Subsequently, we examine the performance of our scheme for key updates in large-scale device scenarios, as outlined in \S~\ref{chp:eva-large-scale} (achieving scalability, flexibility and reliability goal). Finally, we evaluate the overhead on a single device and the agent, discussed in \S~\ref{chp:eva-overhead}.

\subsection{Formal Security Verification}
\label{chp:eva-formal}
ProVerif is an automatic cryptographic protocol verifier, operating in the formal model (also known as the Dolev-Yao model), and is widely used for the formal verification of network protocols. We used it to verify the procedure depicted in Fig.~\ref{fig:get-DS}. The results (List~\ref{lst:result}) confirm that the AK, PK, and communication content are secure and not leaked. The details of the formal verification code written in ProVerif are presented in List~\ref{lst:formal verification}.

\lstset{frame=single, caption={Formal Verification Results.}, label={lst:result}}
\begin{lstlisting}
Verification summary:

Query not attacker(PK[]) is true.

Query not attacker(AK[]) is true.

Query not attacker(successTX[]) is true.
\end{lstlisting}

In our scheme, each device has a unique device key, and the results of the formal security verification demonstrate that our protocol design ensures the security of these keys. We achieved the goal of \textbf{security}.


\subsection{Implementation}
\label{chp:eva-impl}
We develop the agent server prototype, implement the device prototype on the STM32F429I-DISC1 development board, and simulate the usage (update device keys in large scale) of the OTA-Key with Python.

For setup, we use a PC as the agent, which operates on Windows 10 and is equipped with a 12th Gen Intel(R) Core(TM) i5-12490F processor running at 3.00 GHz and 32.0 GB of RAM. We use AES-CBC as the encryption algorithm, chosen for its balance of security and efficiency, providing a reliable encryption mechanism suitable for IoT contexts. 

\textit{Cryptographic Primitives Selected.} In this work, we opt for a symmetric key-based authentication scheme for the following reasons: First, our primary focus is on how to allocate a unique device key to each device before market deployment.
Improving the encryption computation efficiency of existing authentication schemes is beyond the scope of this paper.
Second, considering the resource-constrained nature of IoT devices, the size of certificates and the computational load for communication in the current Public Key Infrastructure (PKI) are greater than those required for symmetric keys. 



\subsection{Impact of Firmware Size on the Key Update Time}
\label{chp:eva-update-time}

\begin{table}
    \centering
    \caption{Typical firmware size. }
    \begin{tabular}{p{4cm} c}
    \hline
        Firmware Type & Size (MB) \\
        \hline
        IoT Gateway (f1) \cite{d-link} & 9.68\\
        DS-2CD2523G0-IS (f2)\cite{hikvision_firmware} & 32.1\\
        DJI Air2 (f3)\cite{DJI}& 175.84\\
        Insta360OneRFW (f4)\cite{Insta360OneRFW}& 76.1\\
        \hline
    \end{tabular}
    \label{tab:firmware-size}
\end{table}

In this subsection, we assess the impact of firmware size on the key update time for a single device across various baselines. We hypothesize a direct connection between the device and the agent server with a network bandwidth of 1MBps.
We select four typical firmware of devices. The devices evaluated include IoT gateways, network surveillance cameras, DJI drones, and gimbal cameras. The firmware sizes for these devices are detailed in Table \ref{tab:firmware-size}. We consider four baselines:

\begin{itemize}
    \item Baseline 1 (BL\_1): BL\_1 represents a traditional firmware update method that does not employ advanced techniques like delta updates~\cite{kim_energy-efficient_2010, kachman_universal_2019}. If the transmission fails or minor differences occur, the entire firmware must be transmitted.

    \item Baseline 2 (BL\_2): BL\_2 incorporates delta download technology~\cite{kim_energy-efficient_2010, kachman_universal_2019}, a well-recognized method to reduce the update size by transmitting only the differences between the new and existing firmware versions. This represents a major improvement over BL\_1, but it requires the entire process to restart upon transmission failure.

    \item Baseline 3 (BL\_3): BL\_3 employs software partitioning techniques~\cite{noauthor_hjsplit_nodate} and resumable downloading techniques~\cite{noauthor_resumable_nodate}. It addresses the limitations of BL\_2 by allowing the retransmission of only the failed segments instead of the entire firmware.

    \item Baseline 4 (BL\_4): For baseline diversity, we also combine the strengths of BL\_2 and BL\_3 as BL\_4. It integrates delta download \cite{kim_energy-efficient_2010, kachman_universal_2019} with software partitioning~\cite{noauthor_hjsplit_nodate} and resumable downloading~\cite{noauthor_resumable_nodate}, representing a state-of-the-art approach in firmware updates.
    
    \item Our approach: We only transmit keys, and if the transmission fails, it starts over from the beginning.
\end{itemize}

\begin{figure}
    \centering
    \includegraphics[width=0.5\textwidth]{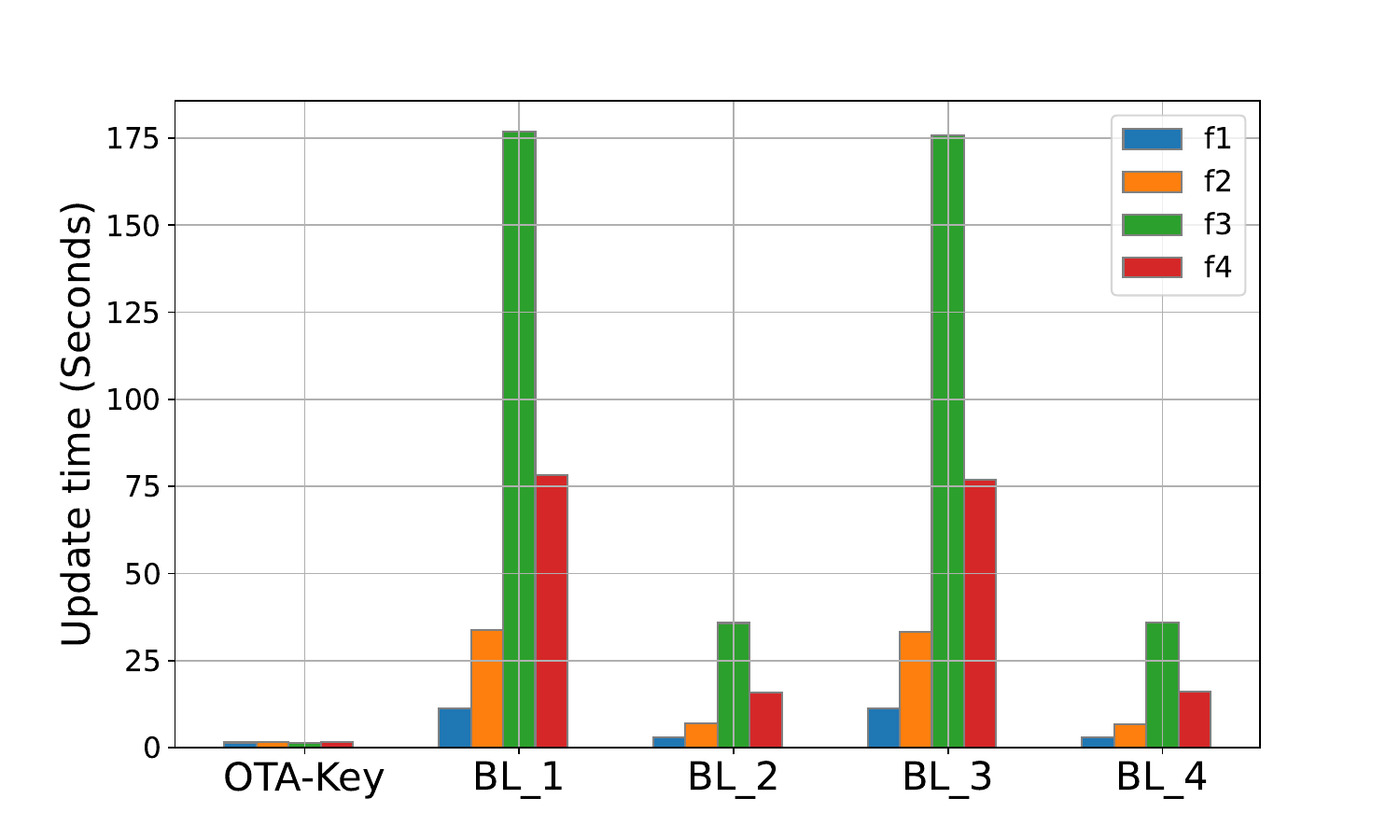}
    \caption{Key update time for a single device with different firmware sizes}
    \label{fig:bar-update-time}
\end{figure}

The results of key update time are shown in Fig.~\ref{fig:bar-update-time}. 
It is evident from the figure that the update time for our OTA-key approach does not vary with firmware size, unlike other baselines where the update time significantly changes based on the firmware size. For example, for firmware around 175MB (indicated by the green bar in the figure), BL\_1 and BL\_3 require approximately 175 seconds; however, for about 10MB of firmware, the update time is only 10 seconds (indicated by the blue bar in the figure). 
The figure also reveals that the update time for BL\_1 and BL\_3 are almost identical, as are those for BL\_2 and BL\_4, which is a reasonable outcome. 
Under this condition, baselines that involve splitting the firmware before transmission show no advantage, hence the update time for `BL\_1 and BL\_3', and `BL\_2 and BL\_4', are nearly the same. BL\_2 and BL\_4, which first analyze the differences between new and old firmware and only transmit these differences, significantly outperform BL\_1 and BL\_3 as they transmit approximately only 20\% of the original firmware data volume \cite{kachman_universal_2019}. 
However, the update time for BL\_2 and BL\_4 is much longer than for our approach. 




%

\subsection{Time and Data Transfer Volume of Key Updates}
\label{chp:eva-large-scale}

In this subsection, we evaluate the time and data transmission volume of our scheme for key updates. We choose the DS-2CD2523G0-IS (f2) \cite{hikvision_firmware} as the firmware for simulation, which has an approximate size of 32.1MB.

We assume a direct connection between the device and the agent, with a total bandwidth of 6.5MBps. 
To simulate real-world conditions, we also include a 5\% failure rate for transmissions, meaning that, on average, updating the keys for 100 devices requires 105 transmissions to successfully complete the task.
We measured the total time and data transfer volume required to complete a single key update for device fleets of 1000, 3000, 5000, 8000, and 10,000 units, respectively. Note that these settings aim to explore the relationship between the key update performance of our scheme with varied device scales. We also analyze whether the scheme supports horizontal scalability to work in large-scale IoT devices in Section~\ref{discussion}.

\begin{figure}
    \centering
    \includegraphics[width=0.5\textwidth]{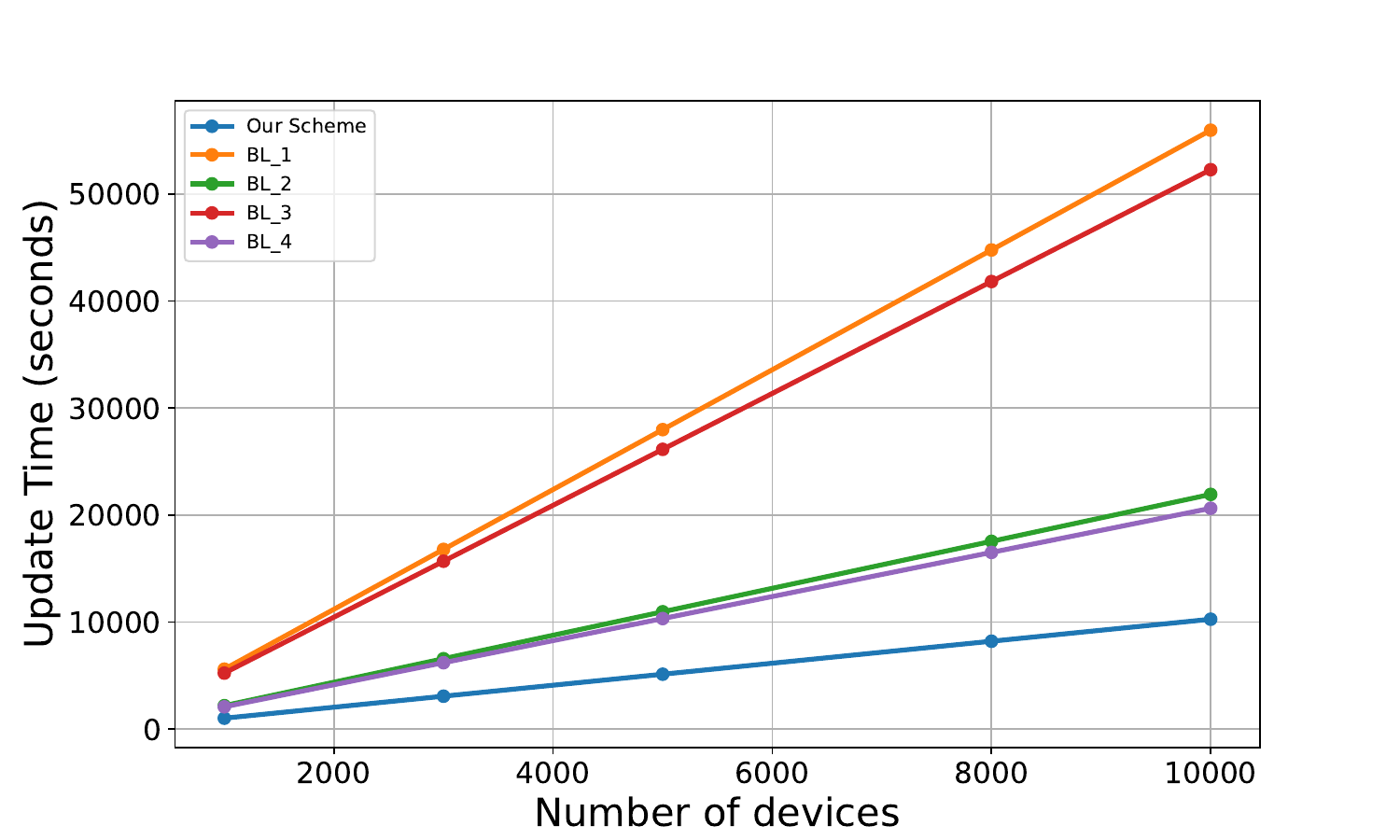}
    \caption{Comparison of time required to complete a single key update for all devices across various baselines.}
    \label{fig:Update time devices}
\end{figure}

\begin{figure}
    \centering
    \includegraphics[width=0.5\textwidth]{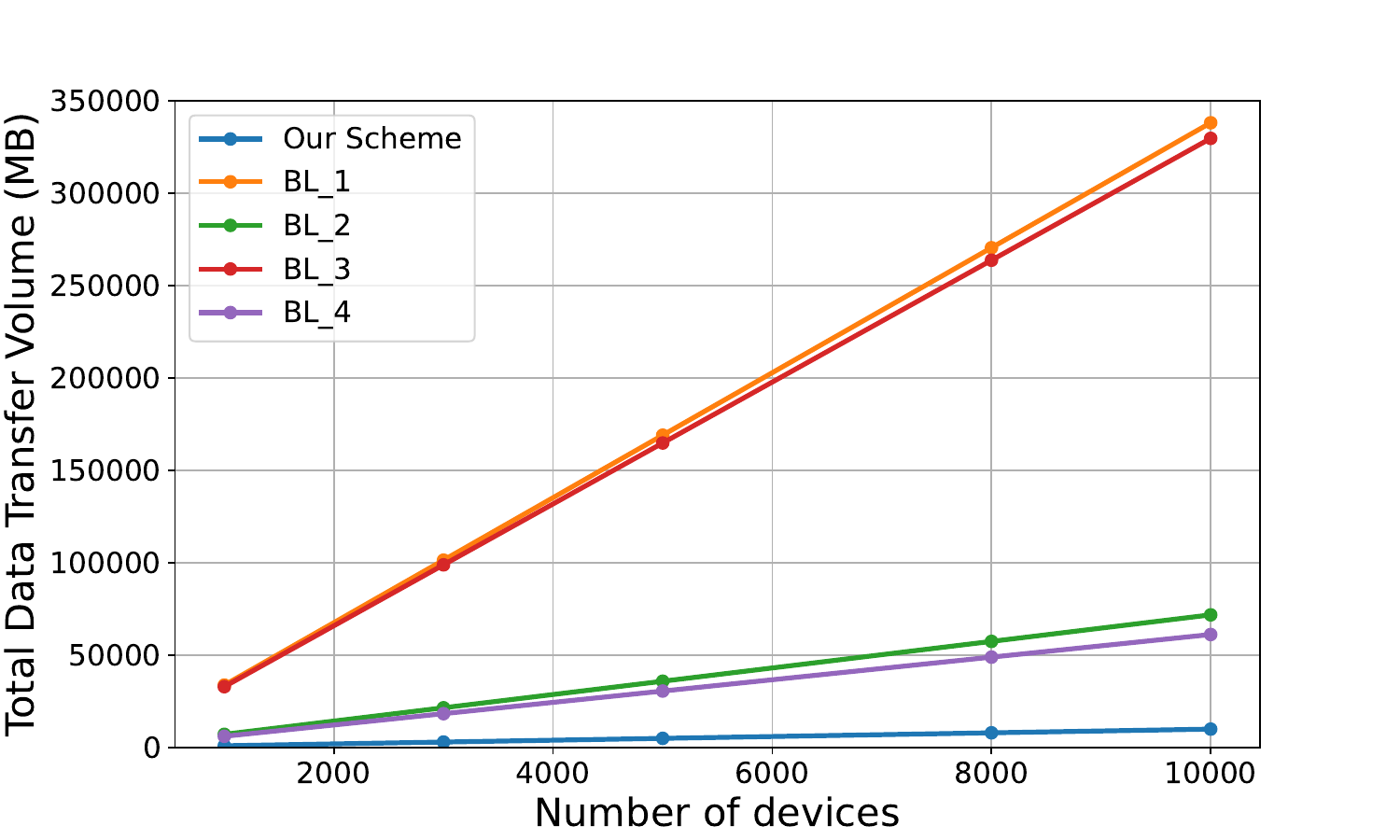}
    \caption{Comparison of data volume required to complete a single key update for all devices across various baselines.}
    \label{fig:Data Transfer Volume}
\end{figure}

Fig.~\ref{fig:Update time devices} illustrates the time required to complete a single key update for all devices, across various baselines and different device scales, with the results presented in seconds.
The figure shows that the update time of our solution is significantly lower than those of other baselines. 

Moreover, Fig.~\ref{fig:Update time devices} shows a linear increase in update time as the number of devices grows, with each baseline distinguished by different colors and styles in the legend. 
From the figure, it is observed that BL\_1 and BL\_3 perform similarly, both showing poor performance with update time exceeding 5000 seconds for 1000 devices. 
In contrast, BL\_2 and BL\_4, which are quite similar in performance, significantly outperform BL\_1 and BL\_3, with update time around 2000 seconds for 1000 devices. 
As previously analyzed, the main difference between BL\_1 and BL\_3 is that BL\_3 does not require restarting the transmission from scratch in case of a failure. 
Therefore, given our overall failure rate is set at a relatively low 5\%, the performance improvement of BL\_3 over BL\_1 is limited. 
BL\_2 and BL\_4 utilize delta download technology, and BL\_4 additionally incorporates the feature of segmenting the transmission content, which results in closely matched performance and overall superiority over BL\_1 and BL\_3.

Fig.~\ref{fig:Data Transfer Volume} displays the total data transfer volumes required to complete a single key update for all devices, across various baselines and different device scales, measured in MB. The figure shows that the data transfer volumes of our solution are significantly lower than those of other baselines. The reasons and analyses for the other results in the figure are similar to Fig.~\ref{fig:Update time devices} previously discussed. Since our evaluation is based on existing production line and the results indicate minimal overhead for key updates, we achieved the goals of \textbf{scalability} and \textbf{flexibility}.

Additionally, we verified that the firmware received by the device matches the original firmware, indicating that our scheme can achieve reliable updates under unstable network conditions. We achieve the goal of \textbf{reliability}.

\subsection{Overhead on a Single Device and the Agent}
\label{chp:eva-overhead}


In this subsection, we use the Power-Z tool to evaluate the impact of our proposed solution on device power consumption. Additionally, we assess the performance of the agent in a large-scale device scenario.

\begin{figure}
    \centering
    \includegraphics[width=0.5\textwidth]{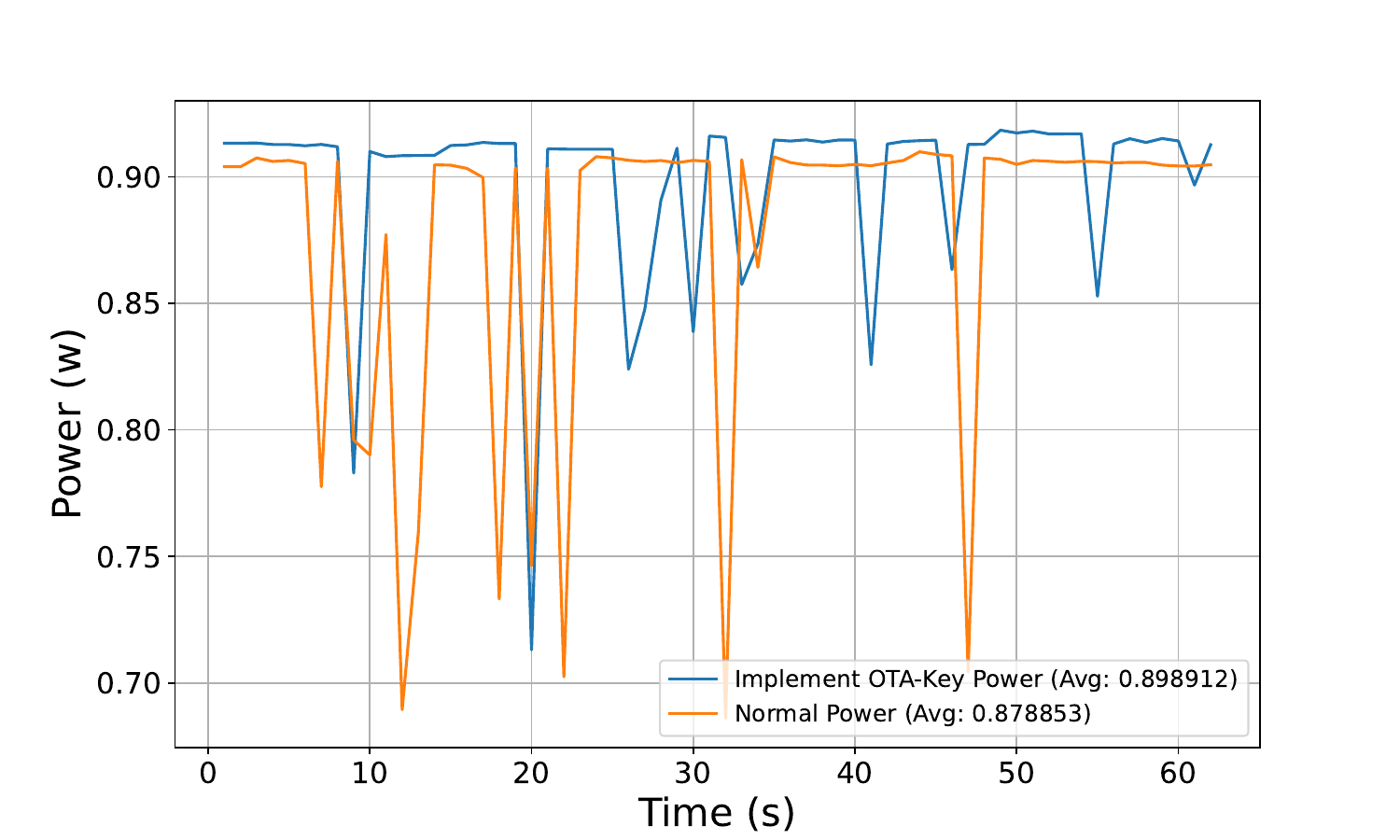}
    \caption{Power processing.}
    \label{fig:power_processing}
\end{figure}

Fig.~\ref{fig:power_processing} compares the power consumption of devices before and after applying the OTA-key solution. The orange curve represents the power consumption after applying OTA-key, while the blue curve shows the original power consumption. The figure shows that the device power consumption does not change significantly after applying our solution. Before the application, the average power consumptionis approximately 0.88W, and after the application, it is around 0.90W, with an increase of no more than 2.3\%. This indicates that our solution does not introduce significant computational overhead, making it suitable for resource-constrained IoT devices.


\begin{table}[ht]
    \centering
    \caption{Performance evaluation of agent server across various device scales.}
    \label{tab:usage}
    \begin{tabular}{c|c|c}
    \hline
    \textbf{Number of Devices}  & \textbf{Memory Usage(MB)} & \textbf{CPU Usage} (\%)\\ \hline
            1000                        & 37.9        &    1.2     \\ \hline
            3000                        & 103.3       &    1.95       \\ \hline
            5000                        & 176.4       &    3.2      \\ \hline
            8000                        & 254.7       &    5.1        \\ \hline
            10000                       & 434.2       &    6.5       \\ \hline
    \end{tabular}
\end{table}

We implement an agent prototype and create multiple clients to simulate real devices communicating with the agent. This setup allows us to evaluate the agent's memory usage as the number of devices increase. Table~\ref{tab:usage} displays the performance of the agent across various device scales. It shows that even with 10,000 devices (according to \cite{velib2024} the number of devices in practice is about 20,000), the agent's memory usage is only 434.2MB, and CPU usage is only 6.5\%, which is acceptable for modern computers.

Overall, the evaluation presented in this subsection demonstrates that our solution does not introduce significant power overhead to the devices nor does it impose a large burden on the server. This makes our solution highly effective for updating keys in resource-constrained IoT devices on a large scale.

\section{Related Work}

We have conducted a comprehensive survey of IoT device provisioning and have also explored related research in the areas of IoT authentication and firmware updating, which are crucial phases in the device provisioning workflow.





\noindent \textbf{Device Provisioning Approaches.} We have reviewed relevant industrial practices, academic literature, and tutorials on various IoT cloud platforms.

\textit{Industrial Practices.} To provision and manage a large number of IoT devices, there are two kinds of methods in the industry. The first is the common practice in the industry \cite{changhong, hisilicon, yiqismart} that inserts one specific authentication key (e.g., a secret key) into the firmware which is further installed on IoT devices with the same model.
Such a method makes lots of IoT devices share the same key, posing severe security threats once one device is compromised. The second is the IoT vendor writes a unique key in the firmware, compiles firmware for each device, and sends the firmware to DM for device installation. 
As stated by Intel\textregistered  SDO \cite{intel_sdo},  the process consumes about approximately 20 minutes for one device and thus is very time-consuming and costly for large-scale device provisioning. 
All of these approaches hardcode cloud-related connection information into the device firmware, failing to support changing cloud platforms. 
Additionally, the Intel\textregistered  SDO only offers a set of standards without providing specific implementations, making it difficult to evaluate. Furthermore, this standard does not emphasize the necessity of assigning unique keys to each device.

\textit{Academic literature.} The approach outlined by Bovskov et al. \cite{bovskov2020time} and Che et al. \cite{che2024blueswat} is limited to Bluetooth devices and does not address the needs of heterogeneous cloud platforms. Suzaki et al. \cite{suzaki2020reboot} focus primarily on rebooting and recovering compromised devices, without considering the distribution of keys to devices. Both approaches assume that the device key is hardcoded into the device firmware, which precludes the independent updating of device keys.

\textit{Tutorials on various IoT cloud platforms.}
We conducted a review of the device provisioning strategies proposed by AWS~\cite{awsDeviceProvision}, Azure~\cite{azureDeviceProvision}, and Alibaba Cloud~\cite{aliDeviceProvision}. AWS's approach focuses solely on how to batch-write device information to the AWS IoT cloud, without addressing how to install keys on the devices. Azure's strategy requires IoT vendors to provide certificate information to the device manufacturer (DM), or allows the DM to directly generate the device key, which could potentially enable the manufacturer to record all device key information. Alibaba Cloud, on the other hand, merely specifies that a unique device certificate must be pre-installed on each device, but does not explain how to implement this step. Additionally, none of these cloud platforms provide specific guidance on how IoT vendors can switch between cloud platforms.



\noindent \textbf{IoT Device Firmware Update Methods.}
\label{subsec:need-firmware-uodate}

Studies by El Jaouhari\cite{el_jaouhari_secure_2022}, Mosenia\cite{mosenia_comprehensive_2017}, and Khan\cite{khan_iot_2018} highlight the vulnerability of IoT devices to various attacks, including those from the Mirai botnet, key exposure leading to privacy leaks, and denial-of-service attacks. These sources underscore the critical need for timely and efficient firmware updates to address security vulnerabilities in IoT devices. Specifically, El Jaouhari\cite{el_jaouhari_secure_2022} points out the security risks stemming from rapid product development that prioritizes market dominance over thorough testing. Existing firmware update solutions fall into two categories: hardware-dependent schemes \cite{akkaoui_resilient_2024}, \cite{asokan_assured_2018}, \cite{xu_dominance_2019} and schemes based on cryptographic primitives \cite{wang_design_2024}, \cite{langiu_upkit_2019}. The first category employs trusted computing technologies, such as Trusted Platform Module (TPM) and Secure Storage, to protect device keys and authenticate firmware sources. However, these solutions require specific hardware and are generally designed for uniform firmware updates across large numbers of devices, which does not cater to the distribution or updating of unique keys for batches of devices. The second category uses cryptographic primitives like signatures and encryption to ensure firmware integrity and confidentiality. However, similar to the first, it primarily supports uniform updates across many devices and is not suitable for distributing or updating unique keys among a large batch of devices.

These schemes do not address the large-scale distribution of unique device keys and do not consider the need for devices to switch cloud platforms.


\noindent \textbf{IoT Authentication Methods.} Currently, IoT devices require keys to authenticate with the cloud and encrypt communications. Although the keys and authentication processes are securely designed, the widespread sharing of the same key among a large number of devices presents significant security risks. This has led to investigations into existing authentication methods. A review of the IoT device authentication literature indicates that most studies focus on communication security during normal use, with limited attention to the distribution of unique device keys and transitions between cloud platforms. For example, Wazid et al. \cite{wazid_secure_2020} and Srinivas et al. \cite{srinivas_anonymous_2020} emphasize communication security but do not address the allocation of unique keys to each device or transitions across cloud platforms. Moreover, recent literature also includes numerous hardware-based authentication schemes, such as hardware fingerprints and Physical Unclonable Functions (PUF), which have been substantial developments. However, most hardware fingerprint schemes \cite{xiao_hardware_2024}, \cite{danev_physical-layer_nodate}, \cite{joo_hold_2020} require specialized hardware, which adds to manufacturing complexity and costs. PUF mechanisms \cite{vaidya_iot-id_2020}, \cite{mall_puf-based_2022}, \cite{nimmy_novel_2023} enhance IoT authentication by dynamically generating cryptographic keys, thereby reducing the risks associated with firmware key storage. Nevertheless, these methods require additional hardware support, extending production cycles, which is disadvantageous for vendors aiming for rapid market penetration. Therefore, our proposed solution does not rely on hardware-based authentication.

\section{DISCUSSION}
\label{discussion}

This section discusses various practical aspects of OTA-Key.

\noindent\textbf{Trust Assumptions of the Agent.} The Agent is securely deployed by trusted IoT vendor, and we assume that attackers cannot compromise the Agent server. 
It is securely deployed and managed by a (trusted) IoT vendor.
To prevent internal attacks, the Agent can be isolated using Trusted Execution Environments (TEEs)~\cite{intelsgx}, which protect sensitive operations and data (e.g., device keys) even from privileged administrators.
To protect against external threats, the Agent can enforce traditional access control techniques \cite{sandhu1998role ,park2002towards}, such as role-based access control, to authenticate entities and restrict unauthorized interactions.

\noindent{\textbf{Prevention Against Product Key Leakage.}} This section primarily discusses measures to prevent the leakage of the product key (PK), the potential consequences of such a leak, and relevant mitigation strategies. As illustrated in Fig.~\ref{fig:system setting}, involving the IoT vendor and Device Manufacturer (DM), we assume that the DM is trusted by the IoT vendor. In the scenario depicted in Fig.~\ref{fig:initial workflow}, the IoT vendor can directly provide the compiled firmware to the DM for debugging purposes. Under these conditions, the PK would not leak unless there is an insider threat within either the DM or IoT vendor.

Even if the PK is compromised, the IoT vendor can detect anomalies during the device initialization phase, as shown in Fig.~\ref{fig:initial workflow}. For instance, the IoT vendor can coordinate with the workers activating the devices to monitor activation times and quantities. If these metrics deviate from the expected values, it could indicate a PK leak. In such cases, the IoT vendor can invalidate the device keys already distributed to the devices and communicate with the cloud to block the affected devices from connecting. Additionally, this scenario exerts real-world pressure on the DM to maintain strict confidentiality of the PK.
Since devices can only connect to the cloud and enter production if they have unique device keys, this scheme achieves the security goal.

\noindent\textbf{Scalability of OTA-Key.} To demonstrate the scalability of the OTA-Key approach, we estimated the time required to assign Unique Agent Keys to devices of varying scales and compared it with current industry practices \cite{changhong, intel_sdo}. The evaluation assumes that all devices have completed initial firmware flashing, with 10 professional engineers on the production line. The estimated time for traditional methods includes firmware compilation and burning (2–5 minutes per device \cite{speed2020}) and manual operations (15–20 minutes per device \cite{intel_sdo}) such as moving devices, connecting them to the workstation, and performing verification. Batch installations can reduce the average time to approximately 2 minutes per device with parallel processing by 10 engineers, but efficiency drops due to repetitive tasks~\cite{rework1, rework2} were considered, leading to a total time of about 2,600 minutes for 1,000 devices. 
The OTA-Key approach includes a key transmission time of approximately 1 second per device, as demonstrated in Fig.~\ref{fig:Update time devices}, and a check time of approximately 3–4 minutes per batch of 100 devices, supported by extensive experimental observations where repeated measurements consistently confirmed this duration.
To mitigate risks of simultaneous updates, the OTA-Key approach adopts a gray release strategy \cite{gray}, performing check every 100 devices. Multiple simulations conducted using a Python script indicated that for 100 devices, the total update time was approximately 5 minutes, comprising 100 seconds for key transmission and 180 seconds for the check. The time required for the OTA-Key approach is approximately one-fortieth of that required by traditional methods to assign unique keys to devices, as shown in Table~\ref{tab:scalability}. These estimates demonstrate that OTA-Key is significantly more efficient for key initialization compared to traditional methods.

\begin{table}[ht]
    \centering
    \caption{Scalability Analysis of OTA-Key and Industry Practices.}
    \label{tab:scalability}
    \begin{tabular}{c|c|c}
        \hline
        \textbf{Number of devices} & \textbf{OTA-Key (min)} & \textbf{Industry Practices (min)}\\
        \hline
        100 & 5 & 200 \\
        \hline
        300 & 16 & 660\\
        \hline
        500 & 27 & 1220\\
        \hline
        1000 & 55 & 2600\\
        \hline
    \end{tabular}
\end{table}

\noindent\textbf{Flexibility of OTA-Key.} During production processes, devices may need to adapt to changes in production plans, necessitating switches between different cloud platforms.
For example, devices may migrate from an on-premises solution to AWS IoT Core \cite{azureDeviceProvision} or Alibaba Cloud\cite{aliyundeviceprovision}.
Besides updating device keys, this may require updating the device firmware because the communication protocols, APIs, and SDKs used by cloud platforms are often platform-specific. A device originally configured for Platform A may not be compatible with Platform B's APIs or protocols without modifications. Updating the firmware ensures the device can support the new platform's requirements, such as MQTT with specific extensions, CoAP, or HTTP over TLS with platform-specific certificates. In this paper, we follow the atomic agent-based device updating method to enable cloud platform switching for IoT devices. First, a uniform firmware update is distributed to devices requiring the platform transition, followed by the distribution of new platform keys via the Agent. Several established solutions \cite{asokan_assured_2018, karthik_uptane_nodate, xu_dominance_2019} already exist to support the implementation of uniform firmware distribution.

\noindent{\textbf{Reliability of OTA-Key.}} As discussed in \S~\ref{sec:atomocUpdate}, OTA-Key ensures the atomicity of device key updates through the ``Atomic Agent-Based Device Updating Method". This approach effectively prevents failures, such as loss of connection to the Agent, caused by unforeseen events like power outages or network interruptions.

\noindent{\textbf{Availability of the Agent.}} In our prototype solution, the Agent serves as an intermediary server to distribute and update unique device keys for the devices.
The availability of the Agent can be achieved using well-established methods~\cite{grayHighAvailiable1991,data_design_2019}.
For example, following the ``Independent failure modes'' design principle outlined in \cite{grayHighAvailiable1991}, the Agent can be deployed across multiple physical nodes based on active-active replication and dynamic load balancing techniques~\cite{data_design_2019}, such that even if one node fails, others can continue to function seamlessly. 
Furthermore, modern cloud platforms, such as Alibaba Cloud \cite{AliyunHA}, AWS \cite{AWSHA1}, and Azure \cite{AzureHA1, AzureHA2}, provide built-in features for high availability, including automatic failover, regional redundancy, and disaster recovery mechanisms. These features can be leveraged to ensure that the Agent remains available under various failure scenarios.

\noindent{\textbf{Horizontal Scalability of the Agent.}} Our scheme supports horizontal scaling, enabling it to handle larger device deployments. Cloud service providers, such as AWS \cite{AWSECS} and Alibaba Cloud \cite{AliECS}, offer mature solutions for horizontal server scaling. These solutions can be leveraged to implement horizontal scaling for the Agent. For instance, the Agent can be expanded to 5 servers, each configured with 10 threads to process device key update requests simultaneously. Under this setup, the theoretical update time for 100,000 devices would be 2,000 seconds (calculated as 10,000 seconds (Fig.~\ref{fig:Update time devices}) *10 /50).

\section{Conclusion}

In this paper, we introduce OTA-Key, a secure, scalable, flexible, and reliable remote device key update scheme designed for distributing and updating unique device keys in large-scale device scenarios. The scheme comprises two main designs: the \textit{two-stage device provisioning strategy} and the \textit{atomic agent-based device updating method}. The two-stage device provisioning strategy decouples device keys from feature firmware and issues keys in two stages, addressing the challenges of mass key distribution without altering the existing production line. The atomic agent-based device updating method updates device keys flexibly via an intermediary server and ensures reliability by reserving a redundant area. We evaluate the performance of the OTA-Key scheme across various device scales, and the results demonstrate its excellence in terms of update time and communication volume.

\ifCLASSOPTIONcaptionsoff
  \newpage
\fi

\bibliographystyle{IEEEtran}
\bibliography{references}

\newpage
\begin{IEEEbiographynophoto}{Qian Zhang} is currently pursuing the M.S. degree with the College of Computer Science and Technology (College of Data Science), Taiyuan University of Technology, China. He received the B.E. degree from Hangzhou Dianzi University, China, in 2020. His research interests include IoT security.
\end{IEEEbiographynophoto}

\vspace{-500pt}

\begin{IEEEbiographynophoto}{Yi He} received the Ph.D. degree from Tsinghua University, Beijing, China. His research interests include system security, in particular, the security of smart devices such as Android and IoT.
\end{IEEEbiographynophoto}

\vspace{-500pt}

\begin{IEEEbiographynophoto}{Yue Xiao} is currently pursuing the M.S. degree with the School of Cyber Science and Engineering.  He  received the B.E. degree from Tsinghua University, China, in 2022. His research interests include IoT and AI security.
\end{IEEEbiographynophoto}

\vspace{-500pt}

\begin{IEEEbiographynophoto}{Xiaoli Zhang} received the Ph.D. degree from Tsinghua University, China, under the supervision of Prof. Jianping Wu in 2020. She is currently an Associate Professor with the School of Computer \& Communication Engineering, University of Science and Technology Beijing, China. Her research interests include trusted computing, verifiable computation, cloud security, and network security.
\end{IEEEbiographynophoto}

\vspace{-500pt}

\begin{IEEEbiographynophoto}{Chunhua Song} is an Associate Professor at Taiyuan University of Technology. She received her Ph.D. degree in 2009 and her M.S. degree in Computer Applications in 2002, both from Taiyuan University of Technology, China. Her research interests include 3D image modeling and visualization, as well as graphics and image retrieval.
\end{IEEEbiographynophoto}

\end{document}